\documentclass[10pt,letterpaper]{article}
\usepackage[top=0.85in,left=2.75in,footskip=0.75in,marginparwidth=2in]{geometry}

\usepackage[utf8]{inputenc}

\usepackage{cite}
\usepackage{tikz}
\usepackage{graphicx}
\usepackage{color}
\usepackage{subfig}
\usepackage{subfloat}
\usepackage{hyperref}
\usepackage{nameref,hyperref}

\usepackage[right]{lineno}

\usepackage{microtype}
\DisableLigatures[f]{encoding = *, family = * }

\raggedright
\usepackage{changepage}

\usepackage[aboveskip=1pt,labelfont=bf,labelsep=period,singlelinecheck=off]{caption}

\makeatletter
\renewcommand{\@biblabel}[1]{\quad#1.}
\makeatother

\usepackage{lastpage,fancyhdr,graphicx}
\usepackage{epstopdf}
\fancyheadoffset[L]{2.25in}
\fancyfootoffset[L]{2.25in}

\usepackage{color}

\definecolor{Gray}{gray}{.25}

\usepackage{graphicx}

\usepackage{sidecap}

\usepackage{wrapfig}
\usepackage[pscoord]{eso-pic}
\usepackage[fulladjust]{marginnote}
\reversemarginpar

\begin{document}
\vspace*{0.35in}

\title{Metadata Enrichment of Multi-Disciplinary Digital Library: A Semantic-based Approach}
\begin{flushleft}
{\Large
\textbf\newline{Metadata Enrichment of Multi-Disciplinary Digital Library: A Semantic-based Approach}
}
\newline
\\
Hussein T. Al-Natsheh \textsuperscript{1,2,3,*},
Lucie Martinet \textsuperscript{4,2},
Fabrice Muhlenbach\textsuperscript{1,5},
Fabien Rico\textsuperscript{1,6},
Djamel A. Zighed\textsuperscript{1,2}
\\
\bigskip
{1} Universit\'e de Lyon, France
\\
{2} Lyon 2, ERIC EA 3083, 5 Avenue Pierre Mend\`es France - F69676 Bron Cedex
\\
{3} CNRS, MSH-LSE USR 2005, 14 avenue Berthelot - F69363 Lyon Cedex 07
\\
{4} CESI EXIA/LINEACT, 19 Avenue Guy de Collongue, F-69130 \'Ecully, France
\\
{5} UJM-Saint-Etienne, CNRS, Lab. Hubert Curien UMR 5516, F-42023 Saint Etienne
\\
{6} Lyon 1, ERIC EA 3083, 5 Avenue Pierre Mend\`es France, F69676 Bron Cedex
\\
\bigskip
* corresponding author: h.natsheh@ciapple.com

\end{flushleft}

\section*{Abstract}
In the scientific digital libraries, some papers from different research communities can be described by community-dependent keywords even if they share a semantically similar topic. 
Articles that are not tagged with enough keyword variations are poorly indexed in any information retrieval system which limits potentially fruitful exchanges between scientific disciplines.
In this paper, we introduce a novel experimentally designed pipeline for multi-label semantic-based tagging developed for open-access metadata digital libraries. 
The approach starts by learning from a standard scientific categorization and a sample of topic tagged articles to find semantically relevant articles and enrich its metadata accordingly. Our proposed pipeline aims to enable researchers reaching articles from various disciplines that tend to use different terminologies. 
It allows retrieving semantically relevant articles given a limited known variation of search terms.
In addition to achieving an accuracy that is higher than an expanded query based method using a topic synonym set extracted from a semantic network, our experiments also show a higher computational scalability versus other comparable techniques. We created a new benchmark extracted from the open-access metadata of a scientific digital library and published it along with the experiment code to allow further research in the topic. 

\smallskip

\noindent{\bf Keywords.} Semantic tagging, Digital~libraries, Topic modeling, Multi-label classification, Metadata enrichment.

\section{Introduction} \label{sec:introduction}

The activity of researchers has been disrupted by ever greater access to online
scientific libraries --in particular due to the presence of open access
digital libraries. Typically when a researcher enters a query for finding
interesting papers into the search engine of such a digital library it is done
with a few keywords. The match between the keywords entered and those used to
describe the relevant scientific documents in these digital libraries
may be limited if the terms used are not the same. Every
researcher belongs to a community with whom she or he shares common knowledge
and vocabulary. However, when the latter wishes to extend the bibliographic
exploration beyond her/his community in order to gather information that
leads him/her to new knowledge, it is necessary to remove several scientific
and technical obstacles like the size of digital libraries, the heterogeneity
of data and the complexity of natural language.

Researchers working in a multi-disciplinary and cross-disciplinary context should have the ability of discovering related interesting articles regardless of the limited keyword variations they know. They are not expected to have a prior knowledge of all vocabulary sets used by all other related scientific disciplines.
Most often, semantic networks \cite{Borgida_Sowa__Principles_of_semantic_networks__1991} are a good answer to the problems of linguistic variations in non-thematic digital libraries by finding synonyms or common lexical fields. 
However, In the scientific research context, using general language semantic network might not be sufficient when it comes to very specific scientific and technical jargons. Such terms also have the challenge of usage evolution over time in which having an updated semantic network counting for new scientific terms would be very expensive to achieve.
Another solution could be brought by the word embedding approach \cite{Mikolov_et_al__Distributed_Representations_of_Words_and_Phrases_and_their_Compositionality__2013}. Another solution could be brought by the word embedding approach \cite{Mikolov_et_al__Distributed_Representations_of_Words_and_Phrases_and_their_Compositionality__2013}.
This technique makes it possible to find semantically similar terms. Nevertheless, this approach presents some problems. It is not obvious to determine the number of terms that must be taken into account to be considered semantically close to the initial term. In addition, this technique does not work well when it comes to a concept composed of several terms rather than a single one. Another strategy is to make a manual enrichment of the digital libraries with metadata in order to facilitate the access to the semantic content of the documents. Such metadata can be other keywords, tags, topic names but there is a lack of a standard taxonomy and they are penalized by the subjectivity of the people involved in this manual annotation process \cite{Abrizah2013}.

In this paper we present an approach combining two different semantic information sources: the first one is provided by the synonym set of a semantic network and the second one from the semantic representation of a vectorial projection of the research articles of the scientific digital library.
The latter takes advantage of learning from already tagged articles to enrich the metadata of other similar articles with relevant predicted tags. 
Our experiments show that the average
F1 measure is increased by 11\% in comparison with a baseline approach that only utilizes semantic networks.
The paper is organized as follows: the next section (Section~\ref{sec:sota})
provides an overview of related work. In Section~\ref{sec:model} we introduce our pipeline of multi-label semantic-based tagging followed by a detailed evaluation in Sections~\ref{sec:expe} and \ref{sec:results}. Finally, Section~\ref{sec:conclusion} concludes the paper and gives an outlook on future work.
\section{State of the Art} \label{sec:sota}

According to the language, a concept can be described by a single term or by an expression composed of multiple words. Therefore the same concept may have different representations in different natural languages or even in the same language in the case of different disciplines. This causes an information retrieval challenge when the researcher does not know all the term variations of the scientific concept he is interested in.
Enriching the metadata of articles with semantically relevant keywords facilitates the access of scientific articles regardless of the search term used in the search engine. Such semantically relevant terms could be extracted thanks to lexical databases (e.g., \textit{WordNet} \cite{miller1995wordnet}) or knowledge bases (e.g.,  \textit{BabelNet} \cite{NavigliPonzetto:12aij},
\textit{DBpedia} \cite{LehmannIJJKMHMK15}, or \textit{YAGO}
\cite{mahdisoltani2014yago3}). Another solution is to use word embedding
techniques \cite{bojanowski2016enriching} for finding semantically similar
terminologies. Nevertheless, it is difficult in this approach to identify
precisely the closeness of the terms in the projection and then if two terms
have still close meanings.

When the set of terms is hierarchically organized, it composes a taxonomy. A
\textit{faceted} or \textit{dynamic taxonomy} is a set of taxonomies, each one
describing the domain of interest from a different point of view
\cite{Sacco_Tzitzikas__Dynamic_taxonomies_and_faceted_search__2009}. Recent
research in this area has shown that it improves the interrogation of
scientific digital libraries to find specific elements, e.g., for finding
chemical substances in pharmaceutical digital libraries
\cite{Wawrzinek_Balke__Semantic_Facettation_in_Pharmaceutical_Collections__2017}.

The use of \textit{Latent Dirichlet Allocation} (LDA) \cite{blei2003latent} for
assigning documents to topics is an interesting strategy in this problem and it
has shown that it helps the search process in scientific digital libraries by
integrating the semantics of topic-specific entities
\cite{Pinto_Balke__Demystifying_the_Semantics_of_Relevant_Objects_in_Scholarly_Collections__2015}.
For prediction problems, the unsupervised approach of LDA has been adapted to a
supervised one by adding an approximate maximum-likelihood procedure to the
process \cite{Blei_McAuliffe__Supervised_Topic_Models__2007}.
Using LDA for topic tagging however has a fundamental challenge in mapping the user defined topics with the LDA's latent topics. We can find a few variations of LDA trying to solve this mapping challenge challenge. For example, \textit{Labeled LDA} technique \cite{Ramage_et_al__Labeled_LDA__2009} is kind of a supervised version of LDA that utilize the user define topic.
Semi-supervised LDA approaches are also
interesting solutions for being able to discover new classes in unlabeled data
in addition to assigning appropriate unlabeled data instances to existing
categories. In particular, we can mention the use of weights of word
distribution in \textit{WWDLDA} \cite{Zhou_et_al__WWDLDA__2013}, or an interval
semi-supervised approach
\cite{Bodrunova_et_al__Interval_Semi-supervised_LDA__2013}.
However, in the case of a real application to millions of documents, such as a
digital library with collections of scientific articles covering many
disciplines, over a large number of years, even recent evolutionary approaches
of LDA require the use of computationally powerful systems, like the use of a
computer cluster \cite{liang15:_large_scale_topic_model}, which is a complex and costly solution.
\section{Model Pipeline}  \label{sec:model}
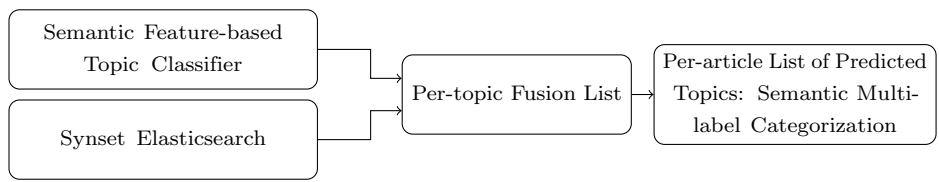
\begin{figure}
\centering
\begin{tikzpicture}
  \node[draw,rounded corners, rectangle, minimum height=3em, text width=11em, text centered] (SFbTC) at (0,3.6) {\footnotesize Semantic Feature-based\\ Topic Classifier};
  \node[draw,rounded corners, rectangle, minimum height=3em, text width=11em, text centered] (SE) at (0,2.4) {\footnotesize Synset Elasticsearch};
  \node[draw,rounded corners, rectangle, minimum height=3em, text width=8em, text centered] (PtFL) at (4.7,3) {\footnotesize Per-topic Fusion List};
  \node[draw,rounded corners, rectangle, minimum height=3em, text width=10em, text centered] (PaLoPT) at (8.4,3) {\footnotesize Per-article List of Predicted Topics: Semantic Multi-label Categorization};
  \coordinate[shift={(0.7,0)}] (A1) at (SFbTC.east);
  \draw[arrows=->] (SFbTC)--(A1)|-(PtFL.172);
  \coordinate[shift={(0.7,0)}] (A2) at (SE.east);
  \draw[arrows=->] (SE)--(A2)|-(PtFL.188);
  \draw[arrows=->] (PtFL)--(PaLoPT);
\end{tikzpicture}
\smallskip
\caption{High-level illustration of the model pipeline. The \textit{Semantic Feature-based Topic Classifier} phase is used to generate \textit{Top N} articles ranked by the probability of topic belonging. Another ranked list is generated by querying the synonym set (synset) of the topic using a text-based search engine which is presented in \textit{Synset Elasticsearch} phase. A \textit{Per-topic Fusion List} is then generated using a special mean rank approach in which only \textit{Top $a \times N$} are considered where $a$ is experimentally determined. Finally, each article is tagged by a list of topics that was categorized with in the \textit{Fusion list}.} \label{fig1}
\end{figure}

The new model we propose can be resumed following a pipeline of 4 main components as illustrated in Figure \ref{fig1}. In this section we will describe each of this components.

\subsection{Semantic Feature-based Topic Classifier} \label{subsec:s3h}
This is computationally a big component that itself includes a pipeline of data transformation and a multi-label classification steps. The main phases of it are described as the following:

\subsubsection{Extract semantic features}
Starting from a multi-disciplinary scientific digital library with an open-access metadata, we extract a big number of articles, i.e., millions in which researchers want to explore. The retrieved data from the metadata of these articles are mainly the \textit{title} and the \textit{abstract}. These two fields will then be concatenated in order to be considered as the textual representation of the article in addition to a unique \textit{identifier}. These set of articles will be denoted as \textit{Corpus}. A TF--IDF weighted bag-of-word vectorization is then applied to transform the \textit{Corpus} into a sparse vector space. This vectorized representation is then semantically transformed into a dense semantic feature vector space, typically 100-600 vector size. The result of this stage is an $(N \times M)$ matrix, where $N$ is the semantic feature vector size and $M$ is the number of articles. It must be accompanied with a dictionary that maps the article unique identifier of the article  to the row index of the matrix.

\subsubsection{Topic classifier}
For each topic name, i.e., scientific category name or a key-phrase of a scientific topic, we generate a \textit{dataset} of \textit{positive} and \textit{negative} examples. The \textit{positive} examples are obtained using a text-based search engine, e.g. \textit{Elasticsearch}, which is a widely used search engine web service built on Apache Lucene, as the resulted articles that have \textit{topic name} matches in \textit{title} OR \textit{abstract}. The negative examples, however, are randomly selected articles from the \textit{Corpus} but with no matches with the \textit{topic name} in any of the metadata text fields. Using this \textit{dataset}, we build a kind of \textit{One-vs-All} topic classifier. This classifier must have the ability of providing the predicted probability value of belonging to the topic, i.e. the class.

\subsubsection{Probability-based multi-label classification}
Each of the obtained \textit{One-vs-All} topic classifiers are then used in a multi-label classification task where each article in \textit{Corpus} will have a probability value of belonging to the topic. This could be thought of as a kind of \textit{fuzzy clustering} or \textit{supervised topic modeling} where the article can be assigned to more than one topic but with a probability of belonging. The result of this stage is a top 100K ranked list of articles per topic with the probability value as the ranking score.

\subsection{Synset Elasticsearch} \label{subsec:synset}
This component is computationally simple but has a great value in the pipeline. It is a kind of query expansion where the query space is increased by finding synonyms and supersets of query terms. So, it also requires a text-based search engine, e.g., \textit{Elasticsearch}. We first need a semantic network or a lexicon database, e.g., WordNet, that can provide a set of synonyms of a giving concept name. For each topic in the set of topics, we generate a set of topic name synonyms, that is denoted by \textit{Synset} (synonym set). Using \textit{Elasticsearch} we then generate a ranked list of articles that have matches in their metadata with any of the synonyms in the topic \textit{Synset}. So, the output of this component is a ranked list of articles per topic. As in Section \ref{subsec:s3h}, this output could be considered as a multi-label classification output but with ranking information rather than a probability score.

\subsection{Fusion and Multi-label Categorization} \label{subsec:fusion}

This final stage constitutes the main contribution part of this experimentally designed pipeline. It uses an introduced ranked list fusion criteria of combining the 2 rankings of an article $A$ which are the rank in the \textit{Synset Elasticseach} list denoted by $s_A$ and the rank in the semantic feature-based topic classifier list, denoted by $r_A$. If an article is present both in the 2 lists, we use a special version of \textit{Mean Rank} score as in Equation \ref{equ:fusion1}. Otherwise, the default score value of the article is given by Equation \ref{equ:fusion2} 
where $|S|$ is the size of the \textit{Synset Elasticseach} list.

\begin{equation}\label{equ:fusion1} t_A=\frac{s_A+r_A}{2} \end{equation}
\begin{equation}\label{equ:fusion2} t_A=r_A \times |S| \end{equation}

The rank score of the \textit{Fusion List} will be finally used to re-rank the articles to generate a new ranked list with a list size that ranges from the $max(|S|, |R|)$ and $|S| + |R|$ where $|R|$ is the size of the semantic feature-based topic classifier list. However, in our model we define a hyper-parameter $a$ that determines the size of the \textit{Fusion} list as in Equation \ref{equ:fusion3}. The hyper-parameter $a$ will be experimentally determined based on multi-label classification statistics and evaluation that would be presented in Section \ref{sec:expe}. 

\begin{equation}\label{equ:fusion3} |F| = a \times |S| \end{equation}

The output of this component, and also the whole pipeline, is a list of articles with their predicted list of topics, i.e. scientific category names. Such list is obtained by applying a \textit{lists inversion} process that takes as input all the per topic \textit{Fusion} lists and generates a per article list of topics for all articles presented in any of the \textit{Fusion} lists.
The obtained list of predicted topics per article are optionally presented with a score value that reflects the ranking of the article in the \textit{Fusion} list of the topic. That score could be used to set an additional hyper-parameter replacing $a$ which would be a score threshold that determines if the topic would be added to the set of predicted topic tags of the article. However, a simple and efficient version, as would be shown in Section \ref{sec:expe}, would only relay of the ranking information but having in place the design parameter $a$.  

\section{Experiments} \label{sec:expe}

\subsection{Data Description}
\subsubsection{Scientific Paper Metadata from ISTEX Digital Library.}
The dataset used for running the experiments is extracted from
\textit{ISTEX}\footnote{Excellence Initiative of Scientific and Technical
Information \href{https://www.istex.fr/}{https://www.istex.fr/}}, a French
open-access metadata scientific digital
library\cite{CNRS__White_Paper_Open_Science_in_a_Digital_Republic__2016}. This
digital library is the result of the \textit{Digital Republic Bill}, a law
project of the French Republic discussed from 2014, one of whose aims is a
``wider data and knowledge
dissemination''\footnote{\href{https://www.republique-numerique.fr/pages/in-english}{https://www.republique-numerique.fr/pages/in-english}}.

ISTEX digital library contains 21 million documents from 21 scientific
literature corpora in all disciplines, more than 9 thousands journals and 300
thousands ebooks published between 1473 and 2015 (in April 2018).

Private publishers (e.g., Wiley, Springer, Elsevier, Emerald...) did not leave
access to their entire catalog of publications, that is why the publication
access does not cover the most recent publications. In addition, because the
contracts were signed with the French Ministry of Higher Education and
Research, even if anybody can access to the general information about the
publications with ISTEX platform (title, names of the authors and full
references of the publication, and also metadata in MODS or JSON format),  the
global access is limited to the French universities, engineering schools, or
public research centers: documents in full text (in PDF, TEI, or plain text
format), XML metadata and other enrichments (e.g., bibliographical references
in TEI format and other useful tools and criteria for automatic indexing).

For our experiments, we considered only a subpart of ISTEX corpus: the articles
must be published during the last twenty years, written in English and related
to sufficient metadata, including their title, abstract, keywords and subjects.

\subsubsection{Scientific Topic from Web of Science}
For each scientific article, we also use a list of tags extracted from the
collection of \textit{Web of
Science}\footnote{\scriptsize{\href{https://images.webofknowledge.com/images/help/WOS/hp_subject_category_terms_tasca.html}{https://images.webofknowledge.com/images/help/WOS/hp\_subject\_category\_terms\_tasca.html}}} which contains more than 250 flattened topics. These
flattened topics are obtained as follows: when a topic is a sub-topic of
another one, we can aggregate to the subcategory terms those of the parent
category (e.g, [computer science, artificial intelligence] or [computer
science, network]). Some of the topics are composition of topics, like ``art
and humanities.''

The selected 33 topics  are: [Artificial Intelligence;  Biomaterials;  biophysics;  Ceramics;  Condensed Matter; Emergency Medicine;  Immunology;  Infectious Diseases;  Information Systems; Literature;  Mechanics;  Microscopy;  Mycology;  Neuroimaging;  Nursing; Oncology;  Ophthalmology;  Pathology;  Pediatrics;  Philosophy;  Physiology; Psychiatry;  Psychology;  Rehabilitation;  Religion;  Respiratory System;  Robotics; Sociology;  Substance Abuse;  Surgery;  Thermodynamics;  Toxicology; Transplantation]

In our experiments, to facilitate the analysis of the results without bias due
to lexical pretreatment, we work only with topics containing neither
punctuation nor linkage words. Moreover, we have kept in our experiences only
\textit{Web of Science} topics with enough articles (in ISTEX digital library)
for having a significant positive subset of documents not used for the learning
part (at least 100 scientific articles). The topics, which can be single words
(as ``thermodynamic'') or a concatenation of words (as ``artificial
intelligence''), should be known in the semantic network to benefit of a
consequent synonyms list. In our work, we present the results obtained with 33 topics, which are English single words or the concatenation of several words.
\subsubsection{Synonym Sets from BabelNet.}

In our experiments, we produce a semantic enrichment by using a list of
synonyms for each concept, also known as ``synset'' (for ``synonym set''). To
build our \textit{synset} list, we need a semantic network. After some
preliminary tests on several semantic networks, we chose \textit{BalbelNet}
 \cite{NavigliPonzetto:12aij} \color{black} which gave better
results. A sample synset from \textit{BabelNet} for the topic \textit{Mycology} is [Mycology, fungology, History of mycology, Micology, Mycological, Mycologists, Study of fungi].

\subsubsection{Supervised LDA}
Based on the state-of-the-art review as described in Section \ref{sec:sota}, we started by developing a model based on LDA. We defined a supervised version of the LDA (\textit{sLDA}) where we the number of topics was set to 33 topics. Each topic was guided by boosting the terms of the topic synonym set obtained from \textit{BabelNet} where the boosting values were [1, 10, 20, 30]. The dataset for experimenting this model were extracted from ISTEX scientific corpus by using \textit{Elasticsearch} getting all articles that have at least one match of any of the 33 topics in any of these metadata fields: \textit{Title}, \textit{abstract}, \textit{subjects} or \textit{keywords}. However, the text used to build the \textit{sLDA} were limited to the \textit{title} and the \textit{abstract}. The evaluation of the \textit{sLDA} model will then be performed on a test set that is constructed from the \textit{keywords} and the \textit{subjects} fields.
\subsection{Experimental Process}\label{sec:exp}
Initially, we defined an accuracy indicator that is based on the count of tagged articles with a list of prediction topics that has at least one label intersection with ground truth. This indicator will be denoted as \textit{At least one common label} metric. The other statistical and multi-label classification evaluation metrics  can be easily found in the literature\footnote{\href{https://en.wikipedia.org/wiki/Multi-label\_classification}{https://en.wikipedia.org/wiki/Multi-label\_classification}}.

In order to build an experiment of our proposed pipeline, we need to experimentally determine some hyper-parameters of it as follows:

\subsubsection{Semantic feature-based topic classifier}
We limit our text representation of the article to its title and abstract, which are available metadata. Comparing Paragraph vector \cite{halko2011finding} and Randomized truncated SVD \cite{halko2011finding} based on a metric that maximizes the inner cosine similarity of articles from the same topics and minimizes it for a randomly selected articles, we choose SVD decomposition of the TF--IDF weighted bag of words and bi-grams resulting in 150 features for more than 4 millions articles. As for the topic classifier, also by comparative evaluation, we select \textit{Random Forest Classifier}, tuning certain design parameters, and use it to rank the scientific corpus. We consider the top 100K articles of each topic classifier to be used in the fusion step.

\subsubsection{Synonym set Elasticsearch}
Reviewing many available semantic networks, we found that BabelNet was the most comprehensive one combining many other networks \cite{NavigliPonzetto:12aij}. So, we use it to extract a set of synonyms, i.e., a \textit{synset} for each topic. This synset is then used to query the search engine of ISTEX which is built on Elasticsearch server. As would be shown in Section \ref{sec:results}. This technique will be used as the experiment baseline.

\subsubsection{Fusion and per multi-label categorization}
The main design parameter of this phase is the size of the ranked list that is achieved by setting it to the double size of the \textit{Synset Elasticsearch} list.

\section{Results and Discussion} \label{sec:results}

First, we run an experiment on \textit{sLDA} as described in Section \ref{sec:expe}. The result of this designed experiment was very disappointing based on the evaluation metrics. The best performing \textit{sLDA} model, that was with a boosting value of 30, resulted in the following evaluation: \textit{F1 measure} = 0.02828, \textit{At-least-one-common-label} =  0.0443, \textit{Jaccard index} =  0.0219 and \textit{Hamming loss} = 0.0798. Comparing to using our pipeline with $a=2$ having \textit{F1 measure} of the 33 topics was 0.6032. So, \textit{sLDA} was obviously not a good candidate to be used as a baseline. However, it was an additional motivation for designing and proposing our pipeline. 
After dropping \textit{sLDA} from further experiments due to the very low evaluation results, we have added 2 more topics to the set of the 33 topics totaling to 35 topics. The 2 additional topics were [International Relations;  Biodiversity Conservation]. We have also added more examples to the test set counting for an additional ISTEX metadata field called \textit{categories:wos} that is actually does not exists in all the articles but was still considered as a good source for increasing the test examples in our published benchmark.

\begin{figure}[htbp]

\centering

\subfloat[Label cardinality *] {\includegraphics[width=.5\linewidth]{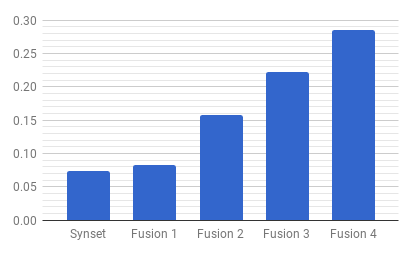}}
\subfloat[Jaccard index **] {\includegraphics[width=.5\linewidth]{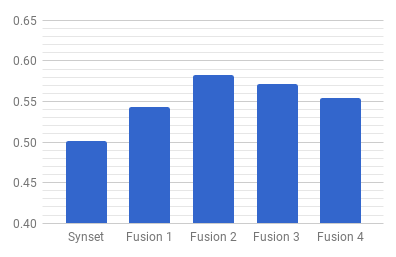}}

\subfloat[Hamming loss $\times 10$] {\includegraphics[width=.5\linewidth]{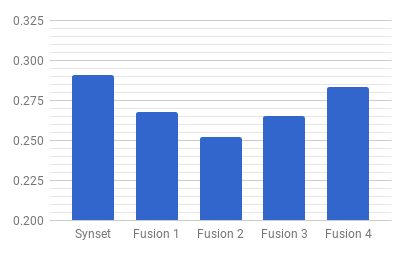}}
\subfloat[F1 measure] {\includegraphics[width=.5\linewidth]{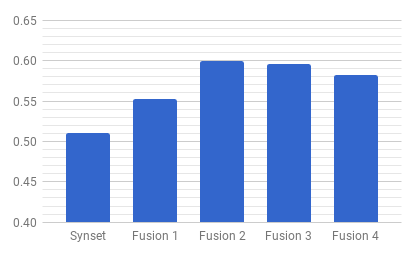}}
\smallskip
\caption{Results of \textit{label cardinality difference}, \textit{Jaccard index}, \textit{Hamming loss} and \textit{F1 measure} evaluation metrics. While Synset is the method that uses synonyms of the category name as a query in Elasticsearch, Fusion 1, 2, 3 and 4 represent respectively the values of the pipeline design parameters $a=[1, 2, 3, 4]$ that determine the number of annotated articles per topic as an integer multiple of the size of \textit{Synset Elasticsearch} list. *: Difference value with the label cardinality of the compared test set of each of the methods.  **: Equivalent to \textit{Precision} in our case of a test set label cardinality = 1.}
\label{fig:eval}
\end{figure}

We define 5 methods for the experiment. One is a method of \textit{Synset Elasticsearch}, denoted here by \textit{Synset} which will be the baseline of benchmark. The other 4 methods are variations of our proposed pipeline but with variant values of the design parameter $a = [1, 2, 3, 4]$. The pipeline methods are then denoted respectively with the value of $a$ as \textit{Fusion1}, \textit{Fusion2}, \textit{Fusion3} and \textit{Fusion4}. The results of the multi-label classification evaluation metrics, described in Section \ref{sec:exp}, are shown in Table \ref{recall} and Figure \ref{fig:eval}. 

\begin{table}
\centering
\caption{
Evaluation results based on the evaluation metrics \textit{Recall} and \textit{At least one common label} denoted here as the \textit{Common-Match} metric. The table also shows the size of the intersection between the method results and the test set that was used in computing the evaluation metric, denoted here as \textit{Intersection}. The value of \textit{Intersection} might also be a good indicator of the method being able to tag more articles.}\label{recall}
\begin{tabular}{|l|r|r| r|}
\hline
Method &  Intersection  & Common-Match & Recall\\
\hline\hline
\textit{Synset}   &  22,192&  0.5284& 0.5285\\
\hline
\textit{Fusion1}  &  22,123&  0.5736& 0.5735\\
\hline
\textit{Fusion2}  &  41,642&  0.6375& 0.6374\\
\hline
\textit{Fusion3}  &  56,114&  \textbf{0.647}0& \textbf{0.6473}\\
\hline
\textit{Fusion4}  &  \textbf{67,625}&  \textbf{0.6470}& 0.6464\\
\hline
\end{tabular}

\end{table}

While the evaluation metric values in Table \ref{recall} recommend  higher $a$ values, 3 or 4 with no significant value difference, we can see from Figure \ref{fig:eval} that the best value is $a=2$ based on \textit{Precision}, \textit{F1 measure}, \textit{Jaccard index }and \textit{Hamming loss}. This means that if we increase the size of the fusion ranked list more than the double of the size of the Synset method, we will start loosing accuracy. Another indicator that we should limit the size of the Fusion list is Figure \ref{fig:eval}.a that shows that if we increase the size of the Fusion list, the difference of the \textit{Label Cardinality} between the predicted results and the compared test set will increase. This difference is a negative effect that should be minimized, otherwise, the model will tend to predict too much labels that would be more probably irrelevant to the article.

Due to the fact that the test set was not generated manually but by filtering on a set of scientific category terms in relevant metadata fields, we believe that  it is an incomplete ground truth. However, we think it is very suitable to compare models as a guidance for designing an efficient one because the test labels are correct even incomplete.
Accordingly, we tried to perform some error analysis where we found that in most of the cases, the extra suggested category names are either actual correct topic having the article a multi-disciplinary one or topics from very similar and related topic. For example, a medical article from ISTEX\footnote{\href{https://api.istex.fr/document/23A2BC6E23BE8DE9971290A5E869F1FA4A5E49E4}{https://api.istex.fr/document/23A2BC6E23BE8DE9971290A5E869F1FA4A5E49E4}} is tagged with the category name [`Transplantation'] in the test set. The predicted topics by our method was [`Mycology', `Transplantation'] resulting into $0.5$ precision value. However, when we read the abstract of that article, we find that it talks about \textit{dematiaceous fungi} which is actually a \textit{Mycology} topic. So, in many cases where there is at least one common tag, the other tags are actually the aimed discovered knowledge rather than a false prediction. The complete list of results --where these cases could be verified-- are published as well as all the experimental data and reproducibility code\footnote{\href{https://github.com/ERICUdL/stst}{https://github.com/ERICUdL/stst}}.

\section{Conclusion and Future Work} \label{sec:conclusion}

Governments, public organizations and even the private sector have recently invested in developing multi-disciplinary open-access scientific digital libraries. However, these huge scientific repositories are facing many information retrieval issues. Nevertheless, this opens opportunities for text-mining based solutions that can automate cognitive efforts in data curation. In this paper, we proposed an efficient and practical pipeline that solves the challenge of the community-dependent tags and the issue caused by aggregating articles from heterogeneous scientific topic ontologies and category names used by different publishers. We believe that providing a solution for such a challenging issue would foster trans-disciplinary research and innovation by enhancing the corpus information retrieval systems. We demonstrated that combining two main semantic information sources --the semantic networks and the semantic features of the text of the article metadata-- was a successful approach for semantic based multi-label categorization. Our proposed pipeline does not only enable for a better trans-disciplinary research but also supports the process of metadata semantic enrichment with relevant scientific categorization tags. 

Other available methods in semantic multi-label categorization, such as LDA, are not suitable in this context for many reasons. For instance, they require powerful computational resources for processing big scientific corpus. Moreover, they need a pre-processing step to detect concepts that are composed of more than one word (e.g., ``Artificial Intelligence''). Finally, LDA is originally an unsupervised machine learning model in which it is problematic to define some undetermined parameters like the number of topics. Our proposed pipeline, however, overcomes all of these limitations and provides efficient results.

Towards improving the query expansion component of the pipeline (Synset Elasticsearch), we are planning to study the impact of using extra information from \textit{BabelNet} semantic network other than only the synonym sets. In particular, we want to include the neighboring concept names as well as the category names of the concept. We expect that such term semantic expansion will improve the performance of the method.


\section*{Acknowledgment}

We would like to thank ISTEX project and ARC6 program\footnote{ \href{http://www.arc6-tic.rhonealpes.fr/larc-6/}{http://www.arc6-tic.rhonealpes.fr/larc-6/}} of the Region Auvergne-Rh\^{o}ne-Alpes that funds the current PhD studies of the first author.

\newpage
\bibliography{library}

\bibliographystyle{abbrv}

\end{document}